# Detecting single infrared photons toward optimal system detection efficiency

PENG HU,[1, 2, 3] HAO LI,[1, 2,3,4] LIXING YOU,[1, 2, 3,5] HEQING WANG,[1, 2, 3] YOU XIAO,[1, 2, 3] JIA HUANG,[1, 2, 3] XIAOYAN YANG,[1, 2, 3] WEIJUN ZHANG,[1, 2, 3] ZHEN WANG,[1, 2,3] AND XIAOMING XIE[1, 2,3]

[1]*State Key Laboratory of Functional Materials for Informatics, Shanghai Institute of Microsystem and Information Technology, Chinese Academy of Sciences(CAS), 865 Changning Rd., Shanghai 200050, China*
[2]*Center of Materials Science and Optoelectronics Engineering, University of Chinese Academy of Sciences, Beijing 100049, China*
[3]*CAS Center for Excellence in Superconducting Electronics, 865 Changning Rd., Shanghai 200050, China*
[4]*lihao@mail.sim.ac.cn;* [5]*lxyou@mail.sim.ac.cn*

**Abstract:** Superconducting nanowire single-photon detector (SNSPD) with near-unity system efficiency is a key enabling, but still elusive technology for numerous quantum fundamental theory verifications and quantum information applications. The key challenge is to have both a near-unity photon-response probability and absorption efficiency simultaneously for the meandered nanowire with a finite filling ratio, which is more crucial for NbN than other superconducting materials (e.g., WSi) with lower transition temperatures. Here, we overcome the above challenge and produce NbN SNSPDs with a record system efficiency by replacing a single-layer nanowire with twin-layer nanowires on a dielectric mirror. The detector at 0.8 K shows a maximal system detection efficiency (SDE) of 98% at 1590 nm and a system efficiency of over 95% in the wavelength range of 1530–1630 nm. Moreover, the detector at 2.1K demonstrates a maximal SDE of 95% at 1550 nm using a compacted two-stage cryocooler. This type of detector also shows the robustness against various parameters, such as the geometrical size of the nanowire, and the spectral bandwidth, enabling a high yield of 73% (36%) with an SDE of >80% (90%) at 2.1K for 45 detectors fabricated in the same run. These SNSPDs made of twin-layer nanowires are of important practical significance for batch production.

## 1. Introduction

Single-photon detection is a key enabling technology applied extensively in modern physics, chemistry, biology, and astronomy. The rapidly advancing quantum information and technology has largely motivated the development of the single-photon detector (SPD) and even require the SPD with near-unity efficiency for numerous quantum fundamental theory verifications [1] and quantum information applications [2-4]. Conventional SPDs, such as avalanche photodiodes and photomultiplier tubes suffer from humble detection efficiencies due to the limited bandgap energy [5]. The recently developed superconducting nanowire single photon detector (SNSPD) promises a high system detection efficiency (SDE) because of the small Cooper-pair breaking energy and is thus widely applied in quantum key distribution [6], optical quantum computing [7], space-ground laser communication [8], satellite laser ranging [9, 10], etc.

 The detection mechanism of SNSPD can be simply modeled as a trigger of a current-carrying superconducting nanowire caused by the hot spot formed after photon absorption [11-14]. As such, the SDE of the SNSPD is primarily determined by the photon-response probability ($\eta_p$) and the optical absorption efficiency ($\eta_{abs}$) of the detector apart from the optical path loss in the system. In previous decades, numerous studies on material selection improvement [15-23], fabrication optimization, and optical structure integration [17, 19, 24, 25] have been performed to improve the SDE in terms of the $\eta_p$ and/or $\eta_{abs}$. To date, SNSPDs using amorphous materials, WSi and MoSi, have demonstrated SDEs of 93%, which however require an ultra-low working temperature of ~100 mK and suffer from low time resolution because of their relatively low superconducting critical temperatures [26].

NbN SNSPDs with higher critical temperatures ($T_c$) showed slightly low SDEs [17, 27-29]. By careful optimization of the geometric parameters and material properties of the nanowires, SDEs of 92% for the 1550-nm wavelength at 2.1K were demonstrated for NbN SNSPDs with a low yield [23, 30]. The challenge to further improve the SDEs for NbN SNSPDs is to have near-unity $\eta_p$ and $\eta_{abs}$ simultaneously for the nanowire with a finite filling ratio, which is limited by the coupled relations between the $\eta_p$ and $\eta_{abs}$ via the superconducting nanowire [31, 32]. To overcome this issue, optical cavities were commonly used to raise the absorption $\eta_{abs}$ that could be further increased to near-unity with the enhanced narrow-band resonant effect, this however in turn leads to the sacrifice of device robustness on the spectral bandwidth and the fabrication tolerance, making the near-unity $\eta_{abs}$ challenging in practice. In addition, a thick nanowire can be employed for the increased absorption length, while it would result in a high critical temperature and thus a low photon-response probability $\eta_p$ [27, 33, 34].

Instead of the thick nanowire, here the sandwiched twin-layer nanowire structure with a very thin insulating layer in between was adopted [35, 36]. This unique structure allows the decoupled $\eta_p$ and $\eta_{abs}$ that ensures the increased optical absorption $\eta_{abs}$ while retaining the $\eta_p$ due to the electrical insulation. When combined with the lossless dielectric mirror, $\eta_p$ and $\eta_{abs}$ can reach nearly 100% simultaneously while maintaining the robustness against practical detector parameters, such as the geometrical size of the nanowire and the spectral bandwidth, thus enabling practical SNSPDs with an optimal SDE and a high yield. The fabricated SNSPD at 0.8K showed a maximal SDE of 98% at a 1590-nm wavelength, and the system efficiency was over 95% in the wavelength range of 1530–1630 nm with a recovery time of 42 ns and a timing jitter of 66 ps. The detector also showed an SDE of 95% at 1550 nm with a dark count rate (DCR) of 100 Hz at 2.1K in a compact Gifford–McMahon (GM) cryocooler.

## 2. Methods, analysis, and devices

Our detector structure was designed with superconducting nanowires fabricated on a dielectric mirror composed of alternative $SiO_2/Ta_2O_5$ films, as shown in Fig. 1(a). This simple structure with a single layer superconducting nanowire (Fig. 1(b)) demonstrated a SDE of ~90% at approximately 2.1K [30]. One issue to further improve for the SDE is the coupled relation between $\eta_p$ and $\eta_{abs}$. To clarify this, Fig. 1(c) presents the $\eta_p$ (red solid line) and $\eta_{abs}$ (blue solid line) at 1550 nm as a function of the nanowire thickness. The photon response $\eta_p$ was obtained by fitting the following empirical expression: $\eta_p(d) = \frac{\eta_0}{(1+(kd)^n)^2}$, where $\eta_0$ denotes the $\eta_p$ at the plateau of small wavelengths λ, the denominator stands for the power law ($n$) describing the decrease in the $\eta_p$ at a larger thickness $d$, and $k$ represents the superconducting nanowire-related parameters (see supplementary materials). The absorption curve results from the electromagnetic calculation (see supplementary materials). Note that the $\eta_p$ saturates to near unity with a nanowire thickness of less than 6 nm and drops dramatically with increased nanowire thickness. This tendency can be easily understood owing to the fact that a limited hot spot formed after the light absorption is not able to efficiently trigger the nanowire with a large cross-section [33, 34, 37]. As opposed to the response, the optical absorption of the single layer SNSPD $\eta_{abs1}$ (green solid line of Fig. 1(c)) increases with the increased film thickness and reached the maximum value at a thickness of approximately 12 nm, which is beyond the commonly used nanowire thickness regarding the low intrinsic photon response.

Generally, the thicker the nanowire, the higher the photon absorption owing to the increased absorption length. However, the thicker the nanowire, the lower the photon response capability. The unmatched maximum values between $\eta_p$ and $\eta_{abs}$ makes the optimal efficiency challenging, as indicated by the product of $\eta_p$ multiplied by $\eta_{abs}$ ($\eta_p \cdot \eta_{abs1}$, triangle scatters of Fig. 1(c)).

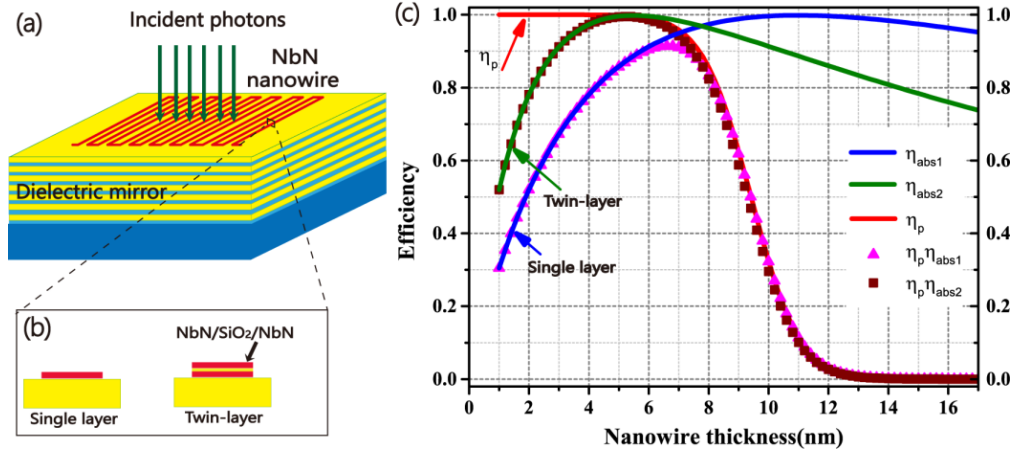

Fig. 1. (a) Schematic of the NbN SNSPD with the superconducting nanowire fabricated on dielectric mirror. The layers of the detector structure, from bottom to top, include the 400-µm-thick Si substrate, dielectric mirror composed of 13 alternative $SiO_2$/$Ta_2O_5$ films, and the NbN superconducting nanowire structure. (b) The cross section of the single layer nanowire structure and the sandwiched twin-layer superconducting nanowire structure defined by the two NbN films with $SiO_2$ films in between. (c) The intrinsic response $\eta_p$, absorption $\eta_{abs1}$/$\eta_{abs2}$, and the product of $\eta_p \cdot \eta_{abs1}$/$\eta_p \cdot \eta_{abs2}$ at 1550 nm as a function of the nanowire thickness for single layer and twin-layer SNSPDs. The $\eta_p$ curve was obtained by fitting the empirical expression and the absorption $\eta_{abs}$ curves were obtained from the electromagnetic calculation (see supplementary materials).

Here, a sandwiched twin-layer superconducting nanowire structure [35, 36, 38] based on a dielectric mirror defined by the two-layer NbN films and $SiO_2$ insulator film between with the same linewidth and the filling factor was employed. With this structure a high $\eta_p$ and a high $\eta_{abs2}$ at a thin thickness of 4−6 nm as shown in Fig. 1(c), could be reached simultaneously. The efficient photon-response capability of each layer nanowire is retained owing to the electrical insulation. When a photon is absorbed by the nanowire in any layer, the generated quasi-particles can pass through the insulator layer and trigger both layers of nanowires almost simultaneously [35]. This trigger regime enables only one channel readout circuit without the after-pulsing issues caused by the electrical instabilities of the conventional avalanched SNSPDs [28, 39]. Thus, this provides a simple and feasible way to decouple the absorption and photon response, promising single photon detection with near-unity efficiency ($\eta_p \cdot \eta_{abs2}$, square scatters of Fig. 1(c)).

Based on the analysis given above, we fabricated SNSPDs with sandwiched twin-layer nanowires, covering a 23-µm-diameter circular area to ensure nearly 100% coupling with the photons illuminated via a single mode fiber. The detector was fabricated on a silicon wafer. The quarter-wave optical film stacks composed of 13 periodic $SiO_2$/$Ta_2O_5$ bilayers with a central wavelength of 1550 nm were alternately deposited onto the Si substrate using ion beam assistant deposition, and the film thickness was optically monitored to ensure adherence to the designed layer thickness. A 6-nm-thick NbN layer was deposited on the substrate at room temperature using reactive DC magnetron sputtering in a mixture of Ar and $N_2$ gases. Then, the thin 3-nm $SiO_2$ was deposited on the NbN film via plasma enhanced chemical vapor deposition. The top 6-nm-thick layer of NbN was deposited via the same process as the bottom NbN film. The sandwiched twin-layer nanowires were obtained via electron beam lithography and reactive ions in the $CF_4$ plasma on the multilayer films. A bridge was finally etched by the reactive ions to form the co-plane waveguide for the readout of electrical signals.

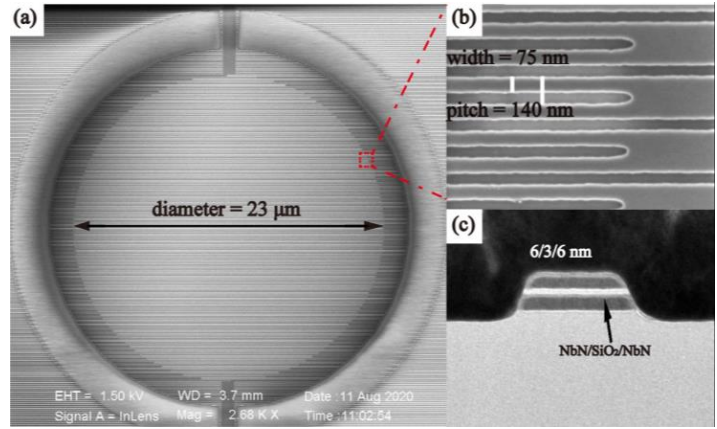

Fig. 2. (a) SEM image of the active area of a sandwiched twin-layer nanowire detector. (b) The fabricated nanowires and the rounded corners of the nanowires were employed to avoid critical current reduction owing to current crowding. (c) The TEM image of the sandwiched twin-layer nanowire structure with each NbN layer at a thickness of 6 nm and insulator $SiO_2$ layer at 3 nm. The device diameter, the nanowire width, and the nanowire pitch were 23 μm, 75 nm, and 140 nm, respectively.

The selected thickness and the width of the superconducting NbN nanowire in each layer were 6 nm and 75 nm, respectively, with a filling factor of $f = 0.54$, which are typical for SNSPDs to respond to near-infrared photons and to ensure a near-unity absorption and considerable fabrication margin. Additionally, optimally rounded boundaries were deployed at the inner corners of the turnarounds to avoid a critical current reduction owing to current crowding [40-42]. Figures 2(a) and (b) show the scanning electron microscopy (SEM) image of a SNSPD with sandwiched twin-layer nanowires, and Fig. 2(c) shows the magnified transmission electron microscopy (TEM) image of the nanowires with the width and pitch of 75 and 140 nm, respectively. The SNSPD has a critical temperature of ~7.3K and a switch current of ~18.3 μA at 0.8K.

## 3. Results and discussion

Figure 3(a) shows the SDE at 1550 nm (red solid squares) and DCR (red dashed squares) of our best detector working at 0.8K provided by a sorption refrigerator. The SDE curve shows a maximal SDE of ≈ 97% and saturated intrinsic photon-response behavior with a large plateau of approximately 3 μA. The spectral SDE provided in Fig. 3(b) shows a maximal SDE of ≈ 98% at 1590 nm and of over 95% from 1530 to 1630 nm. We determined a near-unity broadband efficiency and a large plateau because of the decoupled photon response and photon absorption. The sandwiched twin-layer nanowires enabled the absorption optimization without compromising the photon response. The other lost photons may be attributed mainly to the absorption loss owing to the non-ideal material and fabrication. Additionally, the SDE was measured at 2.1K using a compact GM cryocooler as shown by the blue scatters in Fig. 3(a). This shows a nearly saturated intrinsic photon-response behavior at 1550 nm and a maximal SDE of ≈ 95% at a DCR of 100 Hz.

In our measurement (see supplementary materials), a high precision power meter (Keysight: 81624B; uncertainty of $\sigma_1 = \pm 0.60\%$ calibrated by Physikalisch-Technische Bundesanstalt, PTB) connected to an antireflection coating fiber was used to calibrate the laser power. In the optical link, fiber fusion splicing was used to avoid the optical loss caused by the conventional fiber connector. The uncertainty of the laser output from the optical link caused mainly by the laser fluctuation and attenuation was approximately $\sigma_2 = \pm 0.69\%$ measured using the power meter. Additionally, the uncertainty caused by adjusting the polarization controller was approximately $\sigma_3 = \pm 0.18\%$ and the uncertainty caused by fiber fusion splicing was approximately $\sigma_4 = \pm 0.46\%$ obtained by repeatedly characterizing the loss caused by fusion splicing the fibers. Thus, the total measurement uncertainty of SDE was approximately $\sigma = \pm\sqrt{\sigma_1^2 + \sigma_2^2 + \sigma_3^2 + \sigma_4^2} = \pm 1.0\%$ [26].

We characterized 45 different SNSPDs from the same fabrication batch at approximately 2.1K. In total, 16 (33) detectors with SDEs of over 90% (80%) were obtained (see supporting information for details). The high yield of the high-efficiency SNSPD was ascribed to the detector robustness against the practical parameters, such as the geometrical sizes of the nanowire and quality of the film.

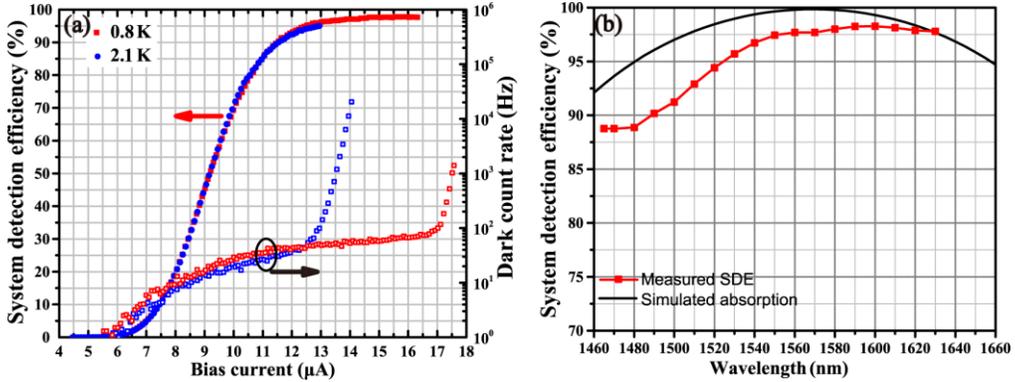

Fig. 3. (a) Bias current dependences of SDE and DCR. The SDE at 1550 nm was characterized at 0.8K (red square) and 2.1K (blue circle), as provided by a sorption refrigerator and a GM cryocooler, respectively. The SDE curve at 0.8K showed an SDE = 97.4 ± 1.0% and a large plateau of ~3 μA at 1550 nm. The SDE curve at 2.1K showed a maximal SDE of 95.0 ± 1.0% at a DCR of 100 Hz and nearly saturated intrinsic photon-response behavior. (b) Wavelength dependences of the SDEs at a bias current of 16.0 μA at 0.8K. The SDE curve shows a maximal SDE of 98.2 ± 1.0% at approximately 1590 nm.

The above results indicated that the sandwiched twin-layer SNSPD provides an effective way to decouple $\eta_{abs}$ and $\eta_p$, thus achieving the near-unity $\eta_p \cdot \eta_{abs}$. Additionally, the twin-layer structure shows improvements in terms of the timing resolution and detection speed because of the smaller kinetic inductance and the higher signal noise ratio compared with the single layer SNSPDs. The timing jitter of the device was measured by recording the histogram of the time difference between the laser-synchronizing signal and the device pulse using a time-correlated single-photon counting module at the wavelength of 1550 nm [43]. Our detector showed a value of 66 ps for the timing jitter, which is defined as the half maximum of the histogram at a bias current of $I_b = 16.0$ μA. At the same bias current, the recovery time constant of the device based on the oscilloscope persistence map was 42 ns, which is defined as the duration of the pulse at 1/e of the maximum pulse amplitude. The detector also showed a counting rate of 20 MHz with half of the maximal efficiency. These results compared with the single-layer detector indicated a faster detection speed and a lower timing jitter (see supporting information).

In principle, a better timing jitter and recovery time can be obtained upon increasing the layer number because of the smaller kinetic inductance and the higher signal noise ratio. Thus, we fabricated and characterized three-layer nanowire SNSPDs. The detector showed a stable detection behavior and an improved recovery time as expected, whereas an unexpected low switching current and thus unsaturated SDE behavior and a low maximum SDE were obtained. We believe some constraints were introduced by the fabrication process. Further optimization of the process is necessary for three or more layered nanowire SNSPDs.

## 4. Conclusions

We demonstrated SNSPDs for optimal system efficiency by fabricating sandwiched twin-layer nanowires on a dielectric mirror. The detector architecture exhibited a decoupled photon response and photon absorption and robustness against the practical parameters of the single-photon detection condition. The best SNSPD at 0.8K showed a maximal SDE of 98% at 1590 nm and an SDE of 95% at 1550 nm in a compact GM cryocooler. A high yield of 73% (36%) with an SDE of >80% (90%) at 2.1K was obtained. We believe it is possible to further improve the yield and push the maximum SDE toward 100%.

### Conflict of interest

The authors declare that they have no conflicts of interest.

### Acknowledgments

We thank Xiaoyu Liu for the help on the electron beam lithography and Photon Technology Co., Ltd for device measurement assistance. This work was supported by National Key R&D Program of China (2017YFA0304000), National Natural Science Foundation of China (61971408, 61671438 and 61827823), Shanghai Municipal Science and Technology Major Project (2019SHZDZX01), Shanghai Rising-Star Program (20QA1410900), Program of Shanghai Academic/Technology Research Leader (18XD1404600), and the Youth Innovation Promotion Association of Chinese Academy of Sciences (2020241).

### Author contributions

H. Li performed the device simulation and designed the experiments. P. Hu fabricated and measured the SNSPDs. H. Wang, Y. Xiao, J. Huang, X. Yang, and W. Zhang provided experimental assistance. H. Li, P. Hu and L. You analyzed the data. All authors contributed to the discussions and the manuscript preparation. H. Li and L. You wrote the manuscript with input from all authors. L. You supervised the project.

# Supplementary information: Detecting single infrared photons toward optimal system detection efficiency

PENG HU,[1, 2, 3] HAO LI,[1, 2,3,4] LIXING YOU,[1, 2, 3,5] HEQING WANG,[1, 2, 3] YOU XIAO,[1, 2, 3] JIA HUANG,[1, 2, 3] XIAOYAN YANG,[1, 2, 3] WEIJUN ZHANG,[1, 2, 3] ZHEN WANG,[1, 2,3] AND XIAOMING XIE[1, 2,3]

[1]*State Key Laboratory of Functional Materials for Informatics, Shanghai Institute of Microsystem and Information Technology, Chinese Academy of Sciences(CAS), 865 Changning Rd., Shanghai 200050, China*
[2]*Center of Materials Science and Optoelectronics Engineering, University of Chinese Academy of Sciences, Beijing 100049, China*
[3]*CAS Center for Excellence in Superconducting Electronics, 865 Changning Rd., Shanghai 200050, China*
[4]*lihao@mail.sim.ac.cn;* [5]*lxyou@mail.sim.ac.cn*

## Table of contents

1. Device simulation

*1.1 Photon-response efficiency, $\eta_p$*

To quantitatively describe the $\eta_p$, we used the following empirical expression of the wavelength-dependent $\eta_p$ [1]: $\eta_p(\lambda) = \frac{\eta_0}{(1+(\frac{\lambda}{\lambda_c})^n)^2}$. Here, $\eta_0$ denotes the $\eta_p$ at the plateau of the spectrum at small wavelengths, whereas the denominator stands for the power law (n) describing the decrease in the $\eta_p$ at larger wavelengths $\lambda$. $\lambda_c$ is the cut-off wavelength. Here, the hot spot model was considered to determine the cut-off wavelength and the $\eta_p$ relations with film thickness [2]:

$$h\frac{c}{\lambda_c} = \frac{N_0 \Delta^2 w d \sqrt{\pi D \tau_{th}}(1-\frac{I_b}{I_c^d})}{\zeta}.$$

Then, the thickness-dependent IDE can be written as

$$\eta_p(d) = \frac{\eta_0}{(1+(kd)^n)^2},$$

where $k = \frac{mN_0 \Delta^2 w d \sqrt{\pi D \tau_{th}}(1-\frac{I_b}{I_c^d})\lambda}{hc\zeta}$. $\Delta$ is the equilibrium superconducting energy gap at the operation temperature. $\zeta$ refers to the quasi-particle multiplication efficiency. $I_c^d$ refers to the depairing critical current [3]. $N_0$ is the density of electronic states. $\tau_{th}$ is the quasi-particle thermalization time. w and d are the linewidth and thickness of the nanowire, respectively. D is the diffusion coefficient. The detector parameters used in our calculation are given as follows. $\zeta = 0.43$ [4] and $\tau_{th} = 7ps$ [5]. $N_0 = \frac{1}{e^2\rho D}$, where $\rho = 300$ $\mu\Omega\cdot cm$ [6], which is the normal-state resistivity [7]. $D = 0.5$ cm$^2$ s$^{-1}$ [8]. $\Delta(0) = 2.0$ k$_B$T$_C$ [8], which is the typical value for 10-nm-thick NbN films. T$_C$ = 7.3 K. $\frac{I_b}{I_c^d} = 0.6$ [9, 10]. To extract the coefficient m and n, we fabricated and characterized single-layer SNSPDs with diameter of 23 µm, linewidth of 75 nm, pitch of 140 nm, and different film thickness. These detectors at 2.1 K showed photon-response efficiencies $\eta_p$ of 1.0, 0.97, 0.79, 0.56, and 0.01 for 5-nm, 6.5-nm, 8-nm, 10-nm, and 12-nm thick SNSPDs. m = 0.51 and n = 10, were then extracted by the fit to these experimental data.

Note that the above analysis is semi-quantitative because the parameters and expression adopted for calculating the $\eta_p$ are either empirical or based on previous publications. The deduced $\eta_p$ based on them may not be precisely accurate. But the method derived from the analysis is correct and a very interesting guide for fabricating SNSPDs with a high detection efficiency.

*1.2 Absorption efficiency*

To quantitatively verify the optical properties of the SNSPDs, the device model was considered to be an infinitely extended periodic structure in the horizontal direction and thus we neglected any edge effects of the real meanders. The incident light was assumed to be a plane wave polarized parallel to the nanowires. The electromagnetic calculation was performed using a rigorous coupled-wave analysis method typically applied for solving scattering problems in periodic structures [11]. The refractive indices we used include 3.46 for Si, 1.44 for SiO$_2$, 2.08 for Ta$_2$O$_5$, and 4.98–4.49i for NbN, which were measured using a spectroscopic ellipsometer at room temperature. The NbN refractive index was obtained based on a 6.5-nm NbN film sample fabricated on a MgF$_2$ substrate. For simplicity, the dispersion effect of the materials in the wavelength range under observation was neglected in our simulation.

We simulated the absorption of SNSPDs with three different structures based on a dielectric mirror: a conventional single layer SNSPD, twin-layer SNSPD, and conventional single layer SNSPD with additional multiple cladding film layers. Our calculations showed the following: (1) the conventional single layer SNSPD requires a thick nanowire that is beyond the commonly used nanowire sizes to achieve near-unity absorption; (2) the bilayer or multilayer SNSPDs show near-unity and robust absorption with commonly used nanowire parameters; and (3) the single layer SNSPD with additional multiple cladding film layers may help to achieve near-unity absorption because of the enhanced resonant effect of the cavity, but sacrifice the robustness of the absorption, such as the bandwidth and the incident angle of the photons. The details of the simulation and results are shown below.

Figure S1 shows the calculated absorption of the single layer SNSPD as a function of the nanowire thickness. We observed that the broadband absorption increased with increased film thickness and reached the maximum value at a thickness of approximately 12 nm.

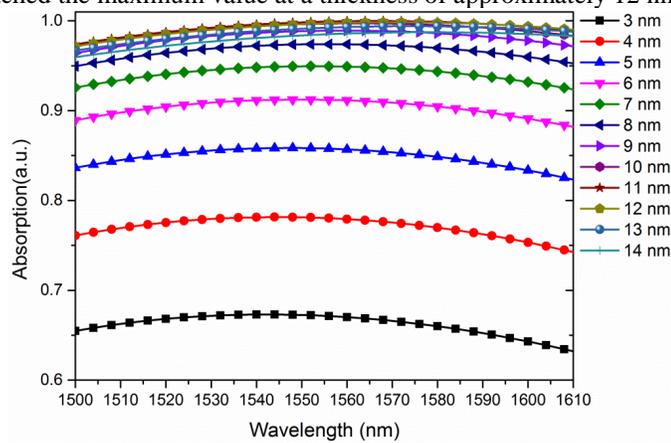

Figure S1. Calculated absorption of single layer SNSPD as a function of nanowire thickness

Figure S2 shows the calculated absorption as a function of light incidence. The calculation shows the robustness against the incident angle of the output light beam from the fiber. The nanoire parameters used in our simulation include a thickness of 6 nm, linewidth of 70 nm, and pitch of 140 nm.

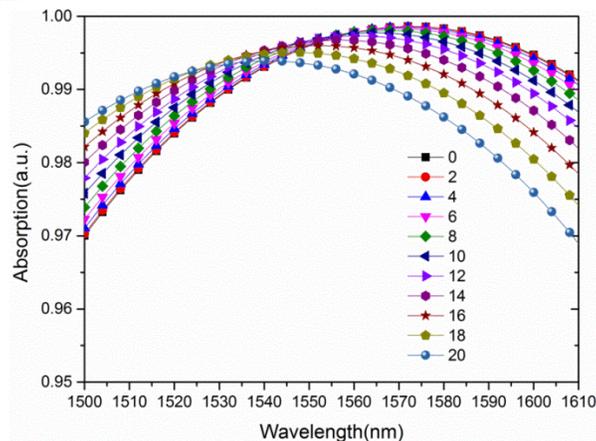

Figure S2. Calculated absorption as a function of light incidence

Figure S3 shows the calculated absorption as a function of the nanowire thickness for single layer SNSPD with another two cladding layers ($SiO_2/Ta_2O_5$) fabricated on top. The calculation shows enhanced and near-unity absorption at approximately the resonant wavelength for the 6-nm-thick nanowires. This can be concluded as the result of the enhanced resonant effect formed by adding another top layers. However, the enhanced resonant effect leads to a lowered robustness of the absorption in terms of the bandwidth as shown in Fig. S3, incident angle of the beam, and geometrical sizes or material properties of the nanowires.

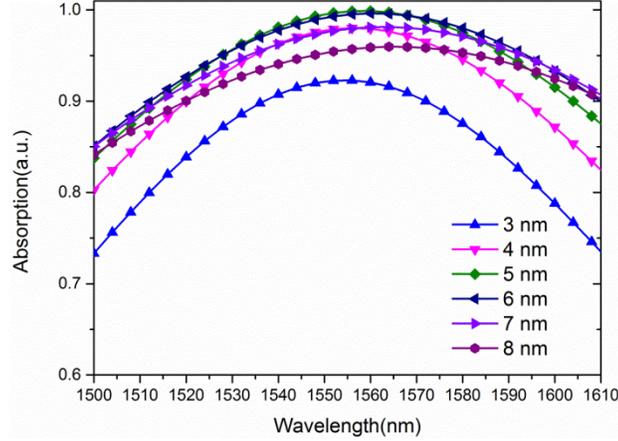

Figure S3. Calculated absorption as a function of the nanowire thickness for single layer SNSPD with another two cladding layers fabricated on top

**2. Device measurement**

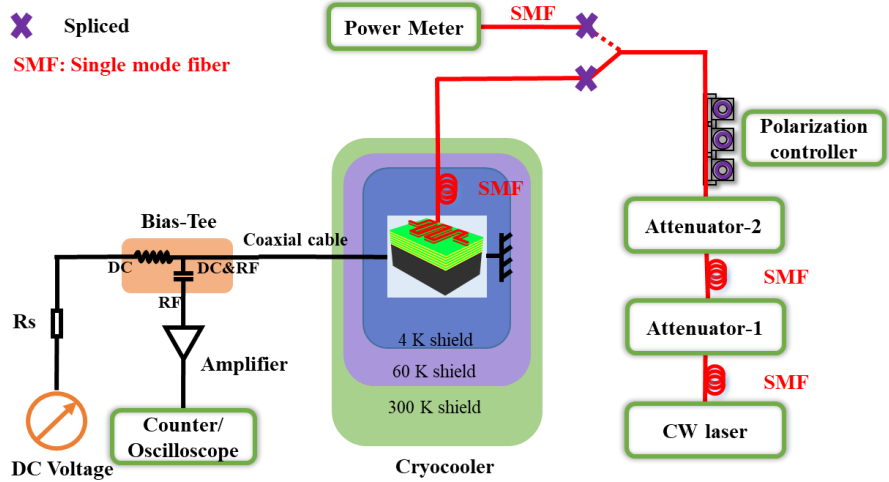

Fig. S4. Device measurement setup. The bias and readout of the SNSPD were connected to the detector through a bias tee and coaxial cables. The bias circuit included an isolated voltage source and a series resistor of 100 kΩ, which formed a quasi-constant current bias to the SNSPD via the DC arm of the bias tee. The voltage pulse generated by the SNSPD was amplified using a room-temperature 50-dB low-noise amplifier (LNA-650, RF Bay Inc.) and then input to a pulse counter or oscilloscope. The fabricated SNSPD was packaged inside a copper sample mount. A single-mode fiber (Corning: SMF-28e) was aligned directly from the front side to the SNSPD. The packaged module was then mounted on the cold head of a two-stage Gifford–McMahon (sorption) cryocooler with a working temperature of 2.1 K (0.85 K). A fiber-coupled continuous wave tunable laser was used in our system detection efficiency (SDE) measurement (Keysight 81970: 1465–1575 nm; Keysight 81940: 1520–1630 nm). The laser was serially connected with two attenuators (Keysight: 81570A) and a polarization controller (Thorlabs: FPC561) to obtain the number of incident photons of ~$10^5$ cps. Fiber fusion splicing was used to avoid the optical loss caused by the conventional fiber connector.

Figure S4 shows the device measurement setup. The fabricated SNSPD was packaged inside a copper sample mount. A single-mode fiber (Corning: SMF-28e) was aligned directly from the front side to the SNSPD. The packaged module was then mounted on the cold head of a two-stage Gifford–McMahon (sorption) cryocooler with a working temperature of 2.1 K (0.8 K). The bias and readout of the SNSPD were connected to the detector through a bias tee and coaxial cables with an impedance of 50 Ω. The bias circuit included an isolated voltage source and a series resistor of 100 kΩ, which formed a quasi-constant current bias to the SNSPD via the DC arm of the bias tee. The voltage pulse generated by the SNSPD was amplified using a room-temperature 50-dB low-noise amplifier (LNA-650, RF Bay, Inc.) with a bandwidth of 30 kHz to 600 MHz and then input to a pulse counter or oscilloscope. A fiber-coupled continuous wave tunable laser was used for our spectral SDEs measurement (Keysight 81970: 1465–1575 nm; Keysight 81940: 1520–1630 nm). The power of the light was attenuated by two attenuators (Keysight: 81570A) to obtain the number of incident photons of ~$10^5$ cps. Additionally, a polarization controller (Thorlabs: FPC561) was used to adjust the light polarization.

In our measurement, a high precision power meter (Keysight: 81624B; uncertainty of $\sigma_1$ = 0.60% calibrated by Physikalisch-Technische Bundesanstalt, PTB) connected to an antireflection coating fiber was used to calibrate the laser power. In the optical link, fiber fusion splicing was used to avoid the optical loss caused by the conventional fiber connector. The uncertainty of the laser power output from the optical link caused mainly by the laser fluctuation and attenuation was approximately $\sigma_2 = \pm 0.69\%$ measured using the power meter. Additionally, the uncertainty caused by adjusting the polarization controller was approximately $\sigma_3 = \pm 0.18\%$ and the uncertainty caused by fiber fusion splicing was approximately $\sigma_4 = \pm 0.46\%$ obtained by repeatedly characterizing the loss caused by fusion splicing the fibers. Thus, the total measurement uncertainty of SDE was approximately $\sigma = \pm\sqrt{\sigma_1^2 + \sigma_2^2 + \sigma_3^2 + \sigma_4^2} = \pm 1.0\%$ [12]. The SDE of the detector is defined as SDE = (PCR − DCR)/PR, where PCR is the output pulse count rate of the SNSPD measured using a pulse counter, DCR is the dark count rate, and PR is the photon rate input to the detection system. At each bias current, an automated shutter in a variable attenuator blocked the laser light, and the dark counts were collected for 10 s. The light was then unblocked, and the pulse counts were measured by collecting pulse counts for another 10 s.

### 3. Sandwiched twin-layer SNSPD

*3.1 RT curve*

Figure S5 presents the resistance as a function of temperature measured in the GM cryocooler, which shows a transition temperature of 7.3 K for our SNSPD.

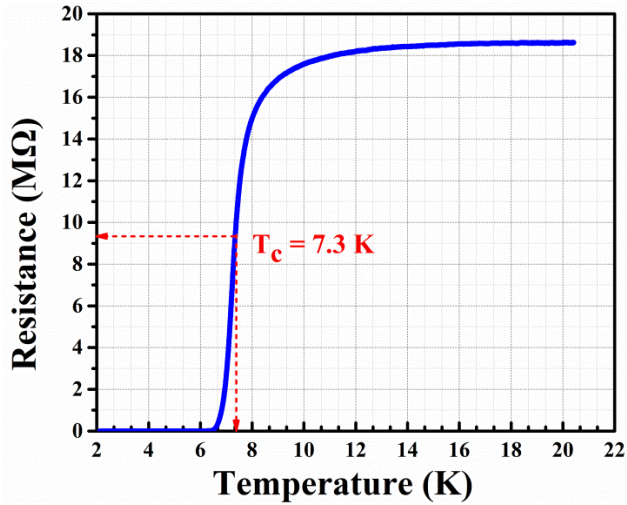

Figure S5. Resistance versus temperature of the sandwiched twin-layer SNSPD

*3.2 Autocorrelation function*

To investigate the operational stability of the device, we measured the autocorrelation function F(τ) of the photon response, which is the correlation of the signal with a delayed copy of itself as a function of the time delay. For detection pulses generated from a continuous wave laser, F(τ) reflects the characteristics of the detector itself. As shown in Fig. S6, a flat F(τ) with a value of 1.0 was obtained at a bias current of 16.0 μA, indicating the absence of after-pulses during the operation of the detector. This means that no electrical avalanche participated in the detection event [13].

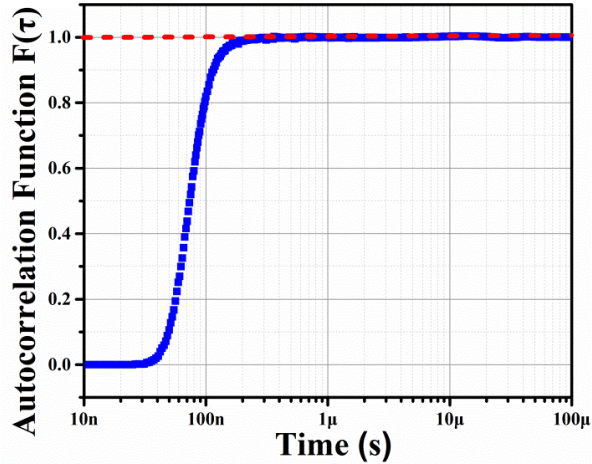

Figure S6. Autocorrelation function F(τ) at a bias current of 16.0 μA

*3.3 Timing jitter*

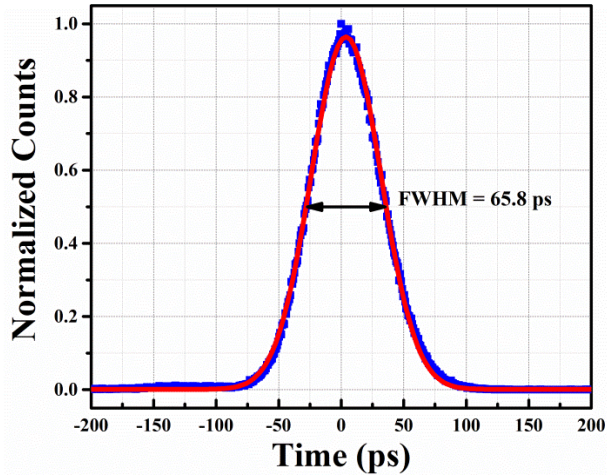

Fig. S7. Histogram of time-correlated photon counts measured at 1550 nm (blue circle). The red solid line is the fitted curve using a Gaussian distribution, whose half maximum of the histogram(FWHM) indicates a system timing jitter of 65.8 ps.

Figure S7 shows the histogram of the time difference between the laser-synchronizing signal and the device pulse. Here, we define the full width of the half maximum of the histogram as the system timing jitter. Our detector shows a value of 65.8 ps at the bias current of $I_b = 16.0$ μA.

*3.4 Recovery time and count rates*

At the same bias current of 16.0 μA, the recovery time τ of the device based on the oscilloscope persistence map in Fig. S8 was 42.0 ns, which is defined as the duration of the pulse at 1/e of the maximum pulse amplitude. The inset of Fig. S8 shows the normalized SDE as a function of count rate, which shows a count rate of 20 MHz with half of the maximal efficiency.

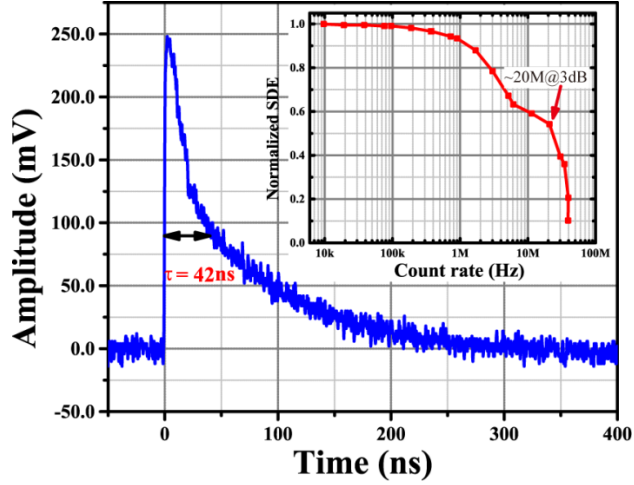

Fig. S8. The oscilloscope persistence map of the response at a bias current of 16.0 μA. The inset shows the normalized SDE as a function of count rate.

## 4. List of the characterized bilayer devices at 1550 nm

Table S1: List of measured devices

| No. | Designed linewidth/pitch (nm) | Resistance at room temperature (MΩ) | Switching current (μA) | SDE@1550 nm | DCR (Hz) | Working temperature (K) |
|---|---|---|---|---|---|---|
| 1 | 80/150 | 13.4 | 12.8 | 88% | ≤100 | 2.1 |
| 2 | 75/140 | 15.1 | 13.1 | 71% | ≤100 | 2.1 |
| 3 | 75/180 | 11.1 | 14.0 | 90% | ≤100 | 2.1 |
| 4 | 85/180 | 9.4 | 17.1 | 91% | ≤100 | 2.1 |
| 5 | 80/150 | 13.8 | 13.0 | 88% | ≤100 | 2.1 |
| 6 | 75/180 | 10.2 | 15.9 | 82% | ≤100 | 2.1 |
| 7 | 85/180 | 8.3 | 15.9 | / | ≤100 | 2.1 |
| 8 | 75/140 | 15.0 | 12.7 | 95% | ≤100 | 2.1 |
| 9 | 75/180 | 10.8 | 4.8 | / | ≤100 | 2.1 |
| 10 | 85/180 | 8.3 | 19.0 | / | ≤100 | 2.1 |
| 11 | 75/180 | 11.0 | 13.9 | 92% | ≤100 | 2.1 |
| 12 | 85/180 | 9.3 | 14.9 | 82% | ≤100 | 2.1 |
| 13 | 80/150 | 12.6 | 15.0 | 87% | ≤100 | 2.1 |
| 14 | 85/180 | 9.3 | 8.8 | 49% | ≤100 | 2.1 |
| 15 | 80/150 | 14.4 | 12.7 | 89% | ≤100 | 2.1 |
| 16 | 75/140 | 15.3 | 13.2 | 90% | ≤100 | 2.1 |
| 17 | 75/180 | 11.2 | 13.7 | 88% | ≤100 | 2.1 |
| 18 | 85/180 | 9.5 | 17.1 | 93% | ≤100 | 2.1 |
| 19 | 80/150 | 12.7 | 13.5 | / | ≤100 | 2.1 |
| 20 | 75/140 | 14.1 | 14.2 | 90% | ≤100 | 2.1 |
| 21 | 85/180 | 9.21 | 16.3 | 84% | ≤100 | 2.1 |
| 22 | 80/150 | 13.8 | 12.2 | 82% | ≤100 | 2.1 |
| 23 | 75/140 | 13.9 | 12.8 | / | ≤100 | 2.1 |
| 24 | 75/180 | 9.5 | 17.6 | 84% | ≤100 | 2.1 |
| 25 | 85/180 | 8.1 | 19.4 | 81% | ≤100 | 2.1 |
| 26 | 75/140 | 15.1 | 13.3 | 88% | ≤100 | 2.1 |
| 27 | 75/180 | 10.1 | 16.6 | 82% | ≤100 | 2.1 |
| 28 | 85/180 | 8.0 | 19.4 | / | ≤100 | 2.1 |
| 29 | 75/180 | 10.7 | 14.9 | 87% | ≤100 | 2.1 |
| 30 | 85/180 | 8.4 | 19.9 | 84% | ≤100 | 2.1 |
| 31 | 80/150 | 11.9 | 16.4 | 88% | ≤100 | 2.1 |
| 32 | 75/180 | 9.8 | 17.4 | 92% | ≤100 | 2.1 |
| 33 | 85/180 | 9.4 | 17.5 | 90% | ≤100 | 2.1 |
| 34 | 80/150 | 13.4 | 14.2 | 89% | ≤100 | 2.1 |
| 35 | 75/140 | 13.8 | 15.2 | 90% | ≤100 | 2.1 |
| 36 | 85/180 | 9.3 | 18.0 | 90% | ≤100 | 2.1 |
| 37 | 75/140 | 9.8 | 11.3 | 72% | ≤100 | 2.3 |
| 38 | 80/150 | 12.3 | 11.7 | 63% | ≤100 | 2.3 |
| 39 | 80/150 | 11.7 | 11.6 | 50% | ≤100 | 2.3 |
| 40 | 80/150 | 12.1 | 15.9 | 90% | ≤100 | 2.1 |
| 41 | 75/140 | 14.1 | 15.5 | 90% | ≤100 | 2.1 |
| | | | 17.1 | 90% | ≤100 | 0.8 |
| 42 | 75/140 | 13.4 | 14.5 | 95% | ≤100 | 2.1 |
| | | | 18.3 | 97% | ≤100 | 0.8 |
| 43 | 75/180 | 10.1 | 15.9 | 92% | ≤100 | 2.1 |
| | | | 18.5 | 95% | ≤100 | 0.8 |
| 44 | 80/150 | 12.0 | 15.9 | 88% | ≤100 | 2.1 |
| | | | 18.8 | 95% | ≤100 | 0.8 |
| 45 | 75/140 | 13.0 | 15.8 | 90% | ≤100 | 2.1 |
| | | | 17.5 | 96% | ≤100 | 0.8 |

Note: / indicates that the detector did not work.

5. **Single layer SNSPD**

For comparision, we presented the timing jitter and the recovery time of single layer SNSPD with diameter of 23 µm, linewidth of 75 nm, pitch of 140 nm. Figure S9 shows the recovery time τ of the device based on the oscilloscope persistence map was 74 ns. Figure S10 shows the timing jitter value of 106.1 ps at the bias current of $I_b = 8.0$ µA.

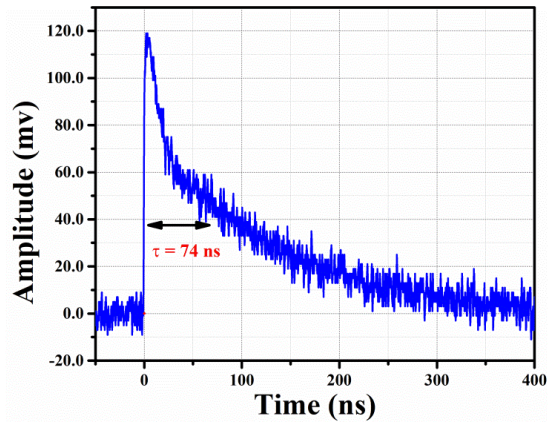

Fig. S9. The oscilloscope persistence map of the response at a bias current of 8.0 μA.

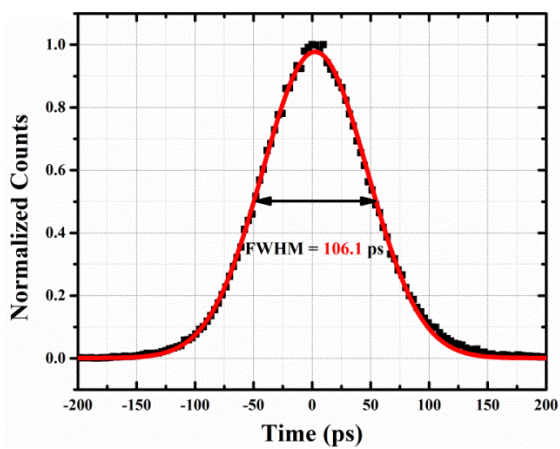

Fig. S10. Histogram of time-correlated photon counts measured at 1550 nm (black square). The red solid line is the fitted curve using a Gaussian distribution, whose FWHM indicates a system timing jitter of 106.1 ps.